\documentclass[a4paper,10pt,twocolumn,nobalancelastpage,aps,pra,superscriptaddress,nofootinbib,longbibliography]{revtex4-2}

\usepackage{libertinus}

\usepackage[utf8]{inputenc}
\usepackage[T1]{fontenc}
\usepackage[kerning=true,protrusion=true,expansion=true]{microtype}

\usepackage{amssymb,amsmath,amsthm,yfonts,pifont,bm,mathrsfs,lettrine,bbold}
\usepackage{longtable,booktabs,tabularx}
\usepackage{siunitx}
\usepackage[cspex,bbgreekl]{mathbbol}
\usepackage{adjustbox}
\DeclareMathAlphabet{\mathbbold}{U}{bbold}{m}{n}

\usepackage{graphicx,color,framed,mdframed,empheq}
\usepackage[dvipsnames,svgnames]{xcolor}
\usepackage{tcolorbox}
\usepackage[scale=0.85,ttdefault=true]{AnonymousPro}
\graphicspath{{./fig/}{./figs/}{./tikz/}}

\usepackage[pdftex,
  colorlinks,
  allcolors=blue,
  pdfauthor={A. Rakhubovsky},
  pdftitle={},
  pdfsubject={},
  pdfkeywords={},
  pdfproducer={Latex with hyperref},
  pdfcreator={pdflatex}]{hyperref}

\usepackage[capitalize,poorman]{cleveref}
\usepackage[mleftright]{diffcoeff}
\usepackage{colonequals}

\newcommand*{\mvec}[1]{\ensuremath{\bm{#1}}}
\newcommand*{\mmat}[1]{\ensuremath{\mathbb{#1}}}

\newcommand*{\Mode}[1]{\ensuremath{\mathsf{#1}}}

\newcommand*{\dg}{\ensuremath{^\dagger}}
\newcommand*{\ee}{\ensuremath{\mathrm{e}}}
\newcommand*{\ii}{\ensuremath{\mathrm{i}}}
\newcommand*{\II}{\mathbb{1}}
\newcommand*{\eye}{\II}
\newcommand*{\qeye}{\II}

\newcommand*{\cT}{\coeff{T}}

\newcommand*{\trp}{\ensuremath{^\intercal}}

\newcommand*{\scrpt}[1]{\mathsf{#1}}
\newcommand*{\s}[1]{\ensuremath{_\scrpt{#1}}}
\newcommand*{\up}[1]{\ensuremath{^\scrpt{#1}}}

\newcommand*{\hc}{\mathrm{h.c.}}

\newcommand{\appropto}{\mathrel{\vcenter{
  \offinterlineskip\halign{\hfil$##$\cr
    \propto\cr\noalign{\kern2pt}\sim\cr\noalign{\kern-2pt}}}}}

\DeclareMathOperator{\cov}{cov}
\DeclareMathOperator{\Tr}{Tr}

\usepackage{mathtools}
\DeclarePairedDelimiter{\bra}{\langle}{\rvert}\DeclarePairedDelimiter{\ket}{\lvert}{\rangle}\DeclarePairedDelimiterX\innerp[2]{\langle}{\rangle}{#1\delimsize\vert\mathopen{}#2}\DeclarePairedDelimiterX\braket[2]{\langle}{\rangle}{#1\delimsize\vert\mathopen{}#2}\DeclarePairedDelimiterX\braketOP[3]{\langle}{\rangle}{#1\,\delimsize\vert\,\mathopen{}#2\,\delimsize\vert\,\mathopen{}#3}\DeclarePairedDelimiterX\ketbra[2]{\lvert}{\rvert}{#1\delimsize\rangle\!\delimsize\langle#2}\DeclarePairedDelimiterX\outerp[2]{\lvert}{\rvert}{#1\delimsize\rangle\!\delimsize\langle#2}\DeclarePairedDelimiterX\projector[1]{\lvert}{\rvert}{#1\delimsize\rangle\!\delimsize\langle#1}\DeclarePairedDelimiterX\dyad[2]{\lvert}{\rvert}{#1\delimsize\rangle\!\delimsize\langle#2}\DeclarePairedDelimiterX\comm[2]{[}{]}{#1,#2}
\DeclarePairedDelimiter{\ev}{\langle}{\rangle}\DeclarePairedDelimiter{\abs}{\lvert}{\rvert}

\newcommand{\mel}[3]{\braketOP{#1}{#2}{#3}}

\colorlet{acolor}{red!75!black}
\colorlet{mcolor}{blue!75!black}
\newcommand*{\angfreq}[2]{2\pi \times \SI{#1}{#2\hertz}}
\newcommand*{\opr}[1]{\hat{#1}}
\newcommand*{\xmm}{\opr X\s m }
\newcommand*{\pmm}{\opr P\s m }
\newcommand*{\xll}{\opr X\s c }
\newcommand*{\pll}{\opr P\s c }

\newcommand*{\rG}{\rho\s{G}}

\newcommand*{\mmC}{\mmat C}

\renewcommand*{\Mode}[1]{\mathscr{#1}}
\newcommand*{\COND}[1]{\mathbb{#1}}
\newcommand*{\CA}{\COND{A}}
\newcommand*{\CS}{\COND{S}}

\begin{document}

\newcommand*{\UPOL}{Department of Optics, Palack{\'y} University, 17.~Listopadu~12, 771~46~Olomouc, Czech~Republic}
\newcommand*{\thetitle}{Quantum Non-Gaussian States of Superfluid Helium Vibrations}
\title{\thetitle}
\author{Andrey A. \surname{Rakhubovsky}}
\email[Corresponding author: ]{rakhubovsky@optics.upol.cz}
\affiliation{\UPOL}
\author{Radim \surname{Filip}}
\email{filip@optics.upol.cz}
\affiliation{\UPOL}

\begin{abstract}
  Quantum non-Gaussian states of phononic systems coupled to light are essential for fundamental studies of single-phonon mechanics and direct applications in quantum technology.
  Although nonclassical mechanical states have already been demonstrated, more challenging quantum non-Gaussianity of such states remains limited.
  Using photon counting detection, we propose the quantum non-Gaussian generation of few-phonon states of low-temperature vibrating superfluid Helium.
  We predict the quantum non-Gaussian depth of such phononic states and investigate their robustness under relevant mechanical heating.
  As the quality of such phononic states is very high, we confirm a single-phonon bunching capability to further classify such states for future mechanical experiments.
  Moreover, we predict increasing capability for force sensing and thermometry for increasing heralded phonon numbers.
\end{abstract}

\date{\today}
\maketitle

\section{Introduction} \label{sec:introduction}

Quantum physics and technology has two paradigmatic directions, with discrete and continuous variables encoding quantum states~\cite{braunstein_quantum_2005,nielsen_quantum_2010,asavanant_optical_2022}.
Each direction has its own advantages and weaknesses, and occasionally there is an effort to combine both to further understand their fundamental interplay and, moreover, to get an advantage for quantum technology~\cite{andersen_hybrid_2015,blais_circuit_2021}.
Quantum cavity optomechanics~\cite{aspelmeyer_cavity_2014,khalili_quantum_2016} has been seemingly leaning mostly towards the continuous-variable domain~\cite{braunstein_quantum_2005} as dictated by the experimental prevalence of linearized optomechanical dynamics and heterodyne detection of light as the main method of the mechanical displacement detection~\cite{clerk_introduction_2010}.
This can be traced from first treatises of ultimate precision of gravity-wave detectors~\cite{braginskii_measurement_1977,caves_measurement_1980,caves_quantummechanical_1981} to modern works~\cite{schnabel_squeezed_2017,mason_continuous_2019,shomroni_optical_2019}.
Experimental studies were gradually focusing on approaching ground states of mechanical oscillators~\cite{teufel_sideband_2011,chan_laser_2011}, and eventually squeezed~\cite{suh_mechanically_2014,wollman_quantum_2015} and entangled states~\cite{palomaki_entangling_2013,ockeloen-korppi_stabilized_2018}.

In recent years, a sizeable effort has been dedicated to develop the control of optomechanical systems at the level of single excitations.
Notable experiments with discrete-variable evaluation of optomechanical two-mode squeezing have been performed~\cite{cohen_phonon_2015,riedinger_nonclassical_2016,hong_hanbury_2017,riedinger_remote_2018}.
There are particularly interesting results coming from adding and subtracting single phonons of mechanical oscillations via counting photons resulting from upconversion of phonons of thermal mechanical vibrations.
Following theoretical proposals~\cite{paternostro_engineering_2011,galland_heralded_2014}, elaborate experiments were able to verify novel out-of-equilibrium character of the resulting states via observing non-Gaussian quadrature distributions~\cite{enzian_singlephonon_2021,patel_roomtemperature_2021} or increased intensity auto-correlation~\cite{patil_measuring_2022a}.
To more precisely characterize the single-phonon control of mechanical motion, further tests are required.

The photon counting detectors possess nonlinearity which is otherwise still challenging to achieve in optomechanical systems at the quantum level.
The detector nonlinearity allows to immediately enter the domain of quantum non-Gaussian optomechanics~\cite{rakhubovsky_quantum_2024}.
Previous notable advances in non-Gaussian mechanical motion were in electromechanics where experiments included coupling mechanics to superconducting discrete-level devices~\cite{oconnell_quantum_2010,chu_creation_2018,clerk_hybrid_2020,chu_perspective_2020,bild_schrodinger_2023,wollack_quantum_2022}.
Non-Gaussianity is an important prerequisite for many tasks in quantum technology as demonstrated by the experiments in other fields~\cite{wolf_motional_2019,mccormick_quantumenhanced_2019,hu_quantum_2019,deng_heisenberglimited_2023,cai_bosonic_2021,dideriksen_roomtemperature_2021,jiang_optically_2023}, is actively pursued experimentally in quantum optomechanics~\cite{jiang_optically_2023,rakhubovsky_quantum_2024}.

Here, inspired by the superfluid Helium setup reported in Refs.~\cite{kashkanova_superfluid_2017,patil_measuring_2022a}, we propose, analyze and confirm the possibility of the conditional quantum non-Gaussian mechanical states, induced by photon-counting detectors.
To achieve conclusive result and comparison with other experiments, we apply criteria~\cite{lachman_faithful_2019,podhora_quantum_2022} of multiphonon and multiphoton quantum non-Gaussianity to these states, and estimate their bunching capability~\cite{zapletal_experimental_2021} and suitability for quantum sensing~\cite{wolf_motional_2019}.
Our results show that the state-of-the-art optomechanical experiment is capable of creating, for the first time, high-quality quantum non-Gaussian states of mechanical motion, which are applicable for quantum sensing.
In near future, these can be used as a phononic resource in quantum error correction~\cite{michael_new_2016}, for quantum sensing of extremely small forces~\cite{lee_using_2019,oh_optical_2020}, or in fundamental tests~\cite{dunningham_nonlocality_2007,kim_scheme_2008,marek_coherentstate_2010} already proposed for photons.
Our main focus is on the superfluid He systems, however, the analysis based on the ubiquitous linearized optomechanical interaction~\cite{law_interaction_1995,bowen_quantum_2015}, and standard theory of photon counting are applicable to other opto-/electromechanical systems.

\section{Results} \label{sec:results}

\begin{figure}[htb]
  \centering
  \includegraphics[width=0.99\linewidth]{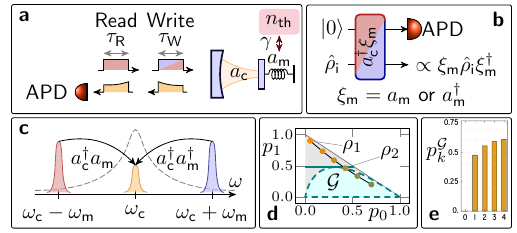}
  \\[1em]
  \begin{tabular}{@{}lllllllllll@{}}
    \multicolumn{11}{c}{Absolute thresholds for $k$-phonon quantum non-Gaussianity}   \\
    \toprule
    $k$                 & 1     & 2     & 3     & 4     & 5     & 6     & 7     & 8     & 9     & 10    \\
    $p_k^{\mathcal{G}}$ & 0.478 & 0.557 & 0.593 & 0.612 & 0.625 & 0.637 & 0.641 & 0.646 & 0.650 & 0.654 \\
    \bottomrule
  \end{tabular}
  \caption{
    Scheme of conditional addition/subtraction of phonons with subsequent verification.
    (a)~A sequence of detuned pulses drives the optomechanical cavity.
    Leaking resonant light is detected by photon-counting detector (APD or a multiplexed version).
    (b)~Cirquit diagram of the process.
    (c)~Frequency-domain illustration of the resonantly enhanced scattering processes.
    Detection of a photon heralds a scattering event which creates/removes a phonon, based on the drive detuning.
    (d,e)~Quantum non-Gaussianity criteria.
    (d)~Example of evaluation. In the plane of probabilities $(p_0, p_1)$ [$p_i = \mel{i}{\rho}{i}$], Gaussian states can only occupy area filled with cyan~($\mathcal G$) bounded by a loss-resilient threshold used for optical states.
    Green horizontal line shows the absolute threshold $p_1^{\mathcal G} = 0.478$ used to quantify mechanical states (see below).
    An initially visibly non-Gaussian state ($\rho_1$, orange dot) gradually thermalizes (darker dots) until it reaches $\rho_2 \in \mathcal G$ and further.
    The admixed noise required to reach $\rho_2$ is the \emph{quantum non-Gaussianity depth} of $\rho_1$.
    (e)~Absolute threshold probabilities $p_k^\mathcal{G}$ for multiphonon quantum non-Gaussianity (see~\cref{sec:phonon_added_thermal_states}) for $0 \leq k < 5$.
    Table below includes values of $p_k^{\mathcal G}$ for larger $k$.
  }
  \label{fig:fig0}
\end{figure}

\subsection{Phonon control of superfluid Helium optomechanics} \label{sec:model}

We consider an optomechanical system formed by mechanical fluctuations of the superfluid Helium in a fiber cavity as reported in~\cite{kashkanova_superfluid_2017,patil_measuring_2022a}.
Such a system exhibits radiation-pressure induced coupling between the optical field and mechanical density waves of the Helium, and in presence of coherent optical driving can be well described by the standard linearized model of optomechanical interaction~\cite{law_interaction_1995,aspelmeyer_cavity_2014} (\cref{fig:fig0}~a).
In particular, driven on the lower mechanical sideband (red detuning, see~\cref{fig:fig0}~c), the system performs excitations swap between the inracavity light and the mechanical oscillator described by the Hamiltonian $H\s{BS} \propto a\s{m} a\s{c}\dg + \hc$,
where $a\s{m} [a\s{c}]$ is the annihilation operator of the mechanics [cavity field].
In the case of driving on the upper sideband (blue detuning), an optomechanical two-mode squeezing interaction (Hamiltonian $H\s{TMS} \propto a\s{m} a\s{c} + \hc$) is realized.
Combining a weak interaction of one of these kinds with a single- or multi-photon detection in the leaking light, it is possible to conditionally approximately add (subtract) a single or multiple phonons to (from) the superfluid Helium~\cite{kashkanova_superfluid_2017,patil_measuring_2022a} (\cref{fig:fig0}~b).
The corresponding conditional quantum state $\rho\s{cond}$ of the mechanics is related to the initial state $\rho_0$ as
\begin{equation}
  \rho\s{cond} = \CA[\rho_0] \appropto { a\s{m}\dg \rho_0 a\s{m} },
  \quad
  \rho\s{cond} = \CS[\rho_0] \appropto { a\s{m} \rho_0 a\s{m}\dg }.
\end{equation}
Due to the technical imperfections in the system (non-unit detection, optical losses, mechanical decoherence, dark counts of the heralding detector, etc.), these states are not exactly single-phonon-subtracted or -added.
A similar technique has been successfully used in experiments with optical photons~\cite{zavatta_quantumtoclassical_2004,zavatta_subtracting_2008}, and atoms~\cite{dideriksen_roomtemperature_2021,corzo_waveguidecoupled_2019}.
For a more formal theoretical discussion, see supplementary materials (SM).
Using sequences of operations $\CA$ and $\CS$, or using photon-number-resolving detectors, it is possible to perform multiple approximate additions and subtractions thereof.
In particular, heralding the mechanical state upon coincidence detections of the output light in a Hanbury-Brown and Twiss (HBT) type of scheme~\cite{yukawa_generating_2013}, performs an approximate addition or subtraction of two phonons at a time.
A similar result can be obtained by repeatedly adding/subtracting a single phonon twice (see comparison in~\cref{sec:phonon_added_thermal_states}).

Here, we consider in detail the protocol to conditionally create a single- and multi-phonon-added mechanical states, and verify their quantum-mechanical properties towards quantum non-Gaussian statistics.
At the heart of this study is the canonical pump-and-probe protocol (\cref{fig:fig0}~(a-c)) that involves a sequence of optical blue-detuned write and red-detuned verification pulses sent to the optomechanical cavity.
All the pulses but the last one are used to perform heralded creation of the conditional mechanical state upon registration of scattered photons by an on-off detector (APD).
The last pulse is red-detuned and is used to swap the conditional quantum state of the mechanical oscillator to the state of the leaking light for the verification.
We consider possible delays between the pulses during which the mechanical oscillator only thermalizes towards equilibrium with the mechanical environment (characterized by damping rate $\gamma$ and mean thermal occupation $n\s{th}$.).
We assume the verification is also performed via multiplexed photon counting (instead of only a single APD), estimate possible visible non-classical effects, and find optimal regimes of operation of the optomechanical protocol.

\subsection{Phonon-added thermal states} \label{sec:phonon_added_thermal_states}

Phonon subtraction and addition are inherently conditional nonlinear operations capable of driving the mechanical oscillator to a non-Gaussian single- or multi-phonon states.
The non-Gaussianity of phonon-number states manifests itself in the shape and negativities of the Wigner function, however, a more detailed evaluation and more precise classification, including depth of quantum non-Gaussianity (see~\cref{fig:fig0}~d), is possible by evaluation of multiphonon contributions of the non-Gaussian states.
Moreover, such evaluation can use direct detection of quantum non-Gaussian features using high-quality single photon counters, as has been performed in quantum optics~\cite{podhora_unconditional_2020}, without need for a complex tomography reconstruction.
Drawing inspiration from Refs.~\cite{lachman_faithful_2019,podhora_quantum_2022}, we compare the phonon-number contributions $\mel{k}{\rho\s{cond}}{k}$ of conditional mechanical state $\rho\s{cond}$ with the thresholds peculiar to multipho(t/n)on quantum non-Gaussian states.
The thresholds (illustrated in~\cref{fig:fig0}~e, also see the table in~\cref{fig:fig0}) show for each  non-negative integer $k$, the maximal probability attainable by states that are the result of Gaussian operations applied to states from the space spanned by all lower Fock states $\ket{i},\ 0 \leq i < k$.
Formally, the threshold probabilities for each $k$ are defined as
\begin{equation}
  \label{eq:absolute:thresh}
  p_k^{\mathcal G} = \max_{\alpha,r, \{c_i\}} \abs{ \braket{k }{  \hat D (\alpha) \hat S (r) \sum_{i = 0}^{k -1} c_i \ket{ i } } }^2.
\end{equation}
where $\hat D (\alpha) = \exp[ \alpha^* a - \alpha a\dg]$ is a Gaussian displacement operator, and $\hat S(r) = \exp[ r (a\dg)^2 - r^* a^2 ]$ is a Gaussian squeezing operator.
The coefficients $c_i$ are arbitrary, but must obey $\sum_i \abs{ c_i }^2 = 1$.
The hierarchy of thresholds~\eqref{eq:absolute:thresh} gradually selects only such approximate mechanical Fock states where adding quantum non-Gaussian features by the detection with more clicks is guaranteed.

\begin{figure}[htb]
  \centering
  \includegraphics[width=0.99\linewidth]{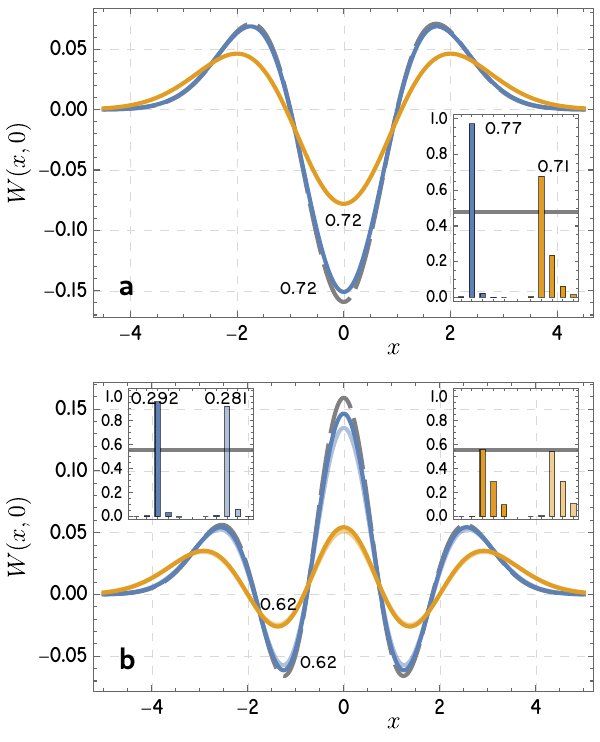}
  \caption{
    Quantum non-Gaussianity of (a)~single- and  (b)~two-phonon-added thermal mechanical state performed via blue-detuned pulses with heralding upon leaking photon detection.
    Single-phonon addition is performed via a single heralding ($\CA$).
    Two-phonon addition in (b) is either via a single pulse with HBT-like coincidence detection for heralding (darker lines), or, alternatively, two subsequent pulses with heralding upon a single APD detection from each pulse (lighter-colored lines).
    Blue lines (bars) correspond to the initial ground state of mechanics ($n_0 = 0$), yellow lines to $n_0 \approx 0.2$, the initial occupation at which the 2-phonon non-Gaussianity of $\CA_2 (\rho_0)$ vanishes.
    Plots in the main panels show the cuts of the Wigner functions of the heralded mechanical states (the Wigner functions are rotationally symmetric).
    Numbers near dips of the Wigner functions show negativity depth (see text).
    Insets show the multiphonon contributions charts for different states.
    Gray horizontal lines indicate thresholds of corresponding multiphonon non-Gaussianity.
    Black numbers show the multiphonon non-Gaussianity depth of the conditional states (number of quanta of thermal noise needed to be added for the corresponding probability to be reduced below the non-Gaussianity threshold).
    }
  \label{fig:non-gauss-2add-pdf}
\end{figure}

In~\cref{fig:non-gauss-2add-pdf} we analyze the non-Gaussianity of the two-phonon-added mechanical states.
Starting with the initial thermal state of the mechanical oscillator characterized by occupation $n_0$, we compare the addition by two subsequent heralding events and by a single heralding via an HBT-like scheme.
We consider a realistic optomechanical interaction described by the superfluid Helium parameters reported in Refs.~\cite{kashkanova_superfluid_2017,patil_measuring_2022a} including thermal decoherence via coupling to environment with mean occupation $n\s{th} = 1.2$.
The mechanical states are evaluated at the instant right after the latest heralding event.

First, we observe that ideally starting from the ground state of the mechanical oscillator, both heralding methods reach a nearly-perfect Fock state ($\ket{1}$ or $\ket{2}$) with the imperfections due to the presence of the mechanical thermal environment.
For two additions, heralding by the HBT detection yields a slightly better result compared to two subsequent single-photon additions due to the shorter overall time of the interaction and consequently a slightly lower influence of the thermal noise.
As a result, the purity and the two-phonon contribution of the HBT-heralded state is higher.
Correspondingly the heralded states from both methods exhibit strong two-phonon non-Gaussianity according to the criteria from~\cite{podhora_quantum_2022}.
We estimate the non-Gaussianity depth of the heralded mechanical state by finding the mean number of thermal phonons to be added to this state to make it lose the two-phonon non-Gaussianity.
To this end, we note that being coupled at the rate $\gamma$ to the thermal bath for the time interval of duration $\tau$, the mechanical oscillator undergoes a transformation that in Heisenberg picture can be written as
\begin{equation}
  a\s{m} (t + \tau) = \ee^{ - \gamma \tau / 2 } a\s{m} (t) + \sqrt{ 1 - \ee^{ - \gamma \tau }} a\s{th},
\end{equation}
where $\gamma$ is the viscous damping rate of the mechanical oscillator, and $a\s{th}$ denotes the effective annihilation operator of a thermal oscillator representing the mechanical environment (see SM for theory details).
Therefore, the number of noise phonons added by this transformation can be expressed
\begin{equation}
  n\s{depth} \approx \tau \gamma n\s{th}.
  \label{eq:depth:define}
\end{equation}
To find the depth of quantum non-Gaussianity or Wigner negativity, we find the duration $\tau$ required for the heralded state to lose the corresponding property due to thermalization, given $\gamma$ and $n\s{th}$ taking the experimental values reported in Refs.~\cite{kashkanova_superfluid_2017,patil_measuring_2022a}, and compute the resulting $n\s{depth}$.

Thermal occupation of the initial mechanical state reduces the quality and the quantum non-Gaussian depth of the heralded states.
This is illustrated by the yellow lines (bars) in~\cref{fig:non-gauss-2add-pdf}.
Initial mechanical occupation $n_0 \approx 0.2$ is the highest that still allows the two-phonon added mechanical state to exhibit the two-phonon QNG (if heralding is by a coincidence HBT-like detection,
the threshold occupation for two subsequent APD heralding is slightly lower).
Notably, the states heralded using either method have visibly non-Gaussian Wigner functions, whose negativity vanishes at much higher initial occupations ($n_0 > 10$).
This illustrates the importance of precise characterization of QNG features of the quantum states under study.

\begin{figure}[htb]
  \centering
  \includegraphics[width=0.99\linewidth]{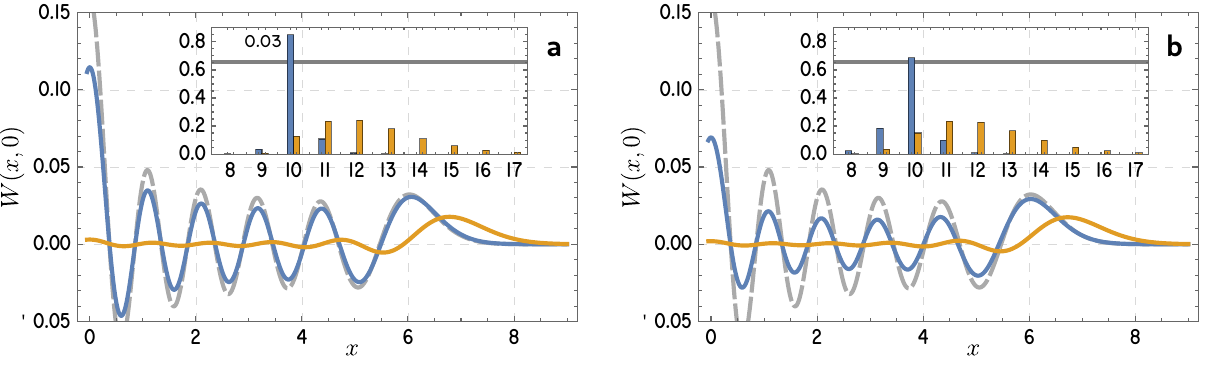}
  \caption{Comparison of the cuts through Wigner functions of the bunched states at the output that approximate the Fock state $\ket{10}$.
  (a) Bunching of $N=10$ copies of single-phonon-added thermal state via HBT.
  (b) $N=5$ copies of two-photons-added thermal state with additional delay nearly sufficient to lose 10-photon non-Gaussianity.
  Without the additional delay, the two panels would be identical.
  Gray dashed line is the theoretical Wigner function of the Fock state $\ket{10}$.
  Insets show multiphonon contributions $[\mel{k}{\rho}{k}]$ of the mechanical states, gray line shows the absolute threshold of $10$-phonon non-Gaussianity $p_{10}^\mathcal{G} = 0.654$, the number near the bar shows the corresponding non-Gaussianity depth (in added noise phonons).
  For blue lines, the initial thermal state is the ground state, for the yellow lines, it is~$\rho\s{th} (0.2)$.
}
  \label{fig:non-gauss-quantif-pdf}
\end{figure}

As the predicted quality of the heralded phonon-number state is high, we investigate the bunching capability of the created non-Gaussian states.
The capability to bunch is highly relevant future linearized optomechanical arrays forming counterparts to trapped ion experiments, applicable in quantum sensing~\cite{gilmore_quantumenhanced_2021} and boson sampling experiments~\cite{chen_scalable_2023}.
To make such a test for future optomechanical experiments, following the approach of Ref.~\cite{zapletal_multicopy_2017,zapletal_experimental_2021} for photons, we consider multiple copies of the conditional mechanical state entering a linear multiport mixing device.
At the output of the mixer we assume an extreme testing event, all phonons conditionally bunching to a single channel, where we compute a function proportional to the Wigner function, in order to evaluate the capability of the output state to exhibit negativities.
The expression for the Wigner function of the output state is in SM.

The simulated Wigner functions are in~\cref{fig:non-gauss-quantif-pdf}.
Multiple copies of heralded mechanical states show the capability to bunch forming higher-order Fock states at the output of a linear quantum network.
Importantly, the bunching capabilities of conditional mechanical states are mostly determined by the thermal noise added during the conditional operation.
In particular, the visibly different conditional states in~\cref{fig:non-gauss-2add-pdf}, show exactly same bunching results (the identical panel for bunching of $N=5$ copies of two-phonon-added state is not shown).
This is because we assumed equal durations of the heralding pulse, and consequently the amount of thermal noise added during the heralding pulse is equal.
Instead, in~\cref{fig:non-gauss-quantif-pdf}~b we show bunching of $N=5$ copies of two-phonon-added state after additional decoherence, sufficient to lose the $10$-phonon quantum non-Gaussianity.

\subsection{Verification of quantum non-Gaussianity} \label{sec:verification_of_quantum_non_gaussianity}

Subsequently, we investigate the possibility to verify the QNG features of the mechanical heralded state by swapping its state to a leaking pulse of light.
The swapping is performed via a red-detuned pulse of classical laser driving which enables a beam-splitter-like interaction \(H\s{BS} \propto a\s{m} a\s{c}\dg + \hc \) in the optomechanical cavity.
The phonons upconverted to photons leak into the detection channel, and can be registered by an optical detector.
From the linear optomechanics theory, we can derive the connection between the quantum state of the mechanics at the initial instant of the readout pulse $t = 0$, and the quantum state of the pulse of light of duration $\tau$ at the detector.
In Heisenberg picture, it reads (see SM for derivation)
\renewcommand*{\cT}{{T}}
\begin{multline}
  \Mode A\s{det} (\tau) = \sqrt{1 - \zeta} \left[ \sqrt{\cT} a\s{m} (0) + \sqrt{1 - \cT} a\s{N} \right]
  \\
  + \sqrt{ \zeta } \: a\s{vac},
  \label{eq:verific:modeoper}
\end{multline}
where $\Mode A\s{det}$ is the annihilation operator of the temporally filtered mode of light (instantaneous field operator $a\s{det} (t)$) with filtering function $f\up{out}(t)$:
\begin{equation}
  \label{eq:verification:modedef}
  \Mode A\s{det} (\tau) = \int_0^\tau a\s{det} (t) f\up{out}(t) \dl{t}.
\end{equation}
Hereinafter we address this mode as the \emph{detector mode}.

In~\cref{eq:verific:modeoper}, $\cT$ is the effective readout transmittance (readout efficiency) which depends on the parameters of the optomechanical interaction (coupling rate, linewidths of cavity and the mechanics) and the choice of the temporal filtering function $f\up{out}(t)$.
In derivation of $\cT$ it is assumed that the optomechanical cavity has only one output channel (no internal losses), and in our simulations we assumed the optimal filtering function $f\up{out}$ that maximizes $\cT$.
The problem of finding temporal modes of flying degrees of freedom has been studied in e.g.~\cite{tufarelli_input_2012,tufarelli_coherently_2014}.
The internal cavity losses, loss in propagation towards the detector, and the detector inefficiency are accounted for in~\cref{eq:verific:modeoper} by the effective transmittance $\zeta$, and the corresponding vacuum noise $a\s{vac}$.

\begin{figure}[htb]
  \centering
  \includegraphics[width=0.99\linewidth]{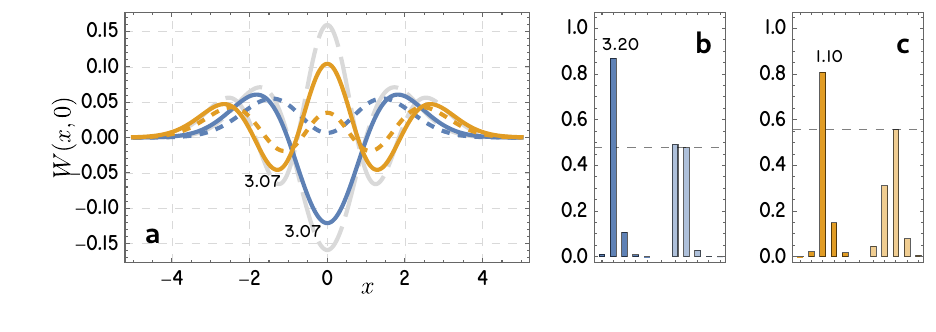}
  \caption{Quantum non-Gaussianity of the states of light at the detector.
    (a) Wigner function cuts of the single-phonon-added (blue lines) and two-phonon-added (yellow lines) thermal states, assuming perfect readout $\zeta = 0$ (full lines) and the value of $\zeta$ at which multiphonon QNG vanishes.
    The corresponding value of $\zeta$ (in units $-10 \log_{10} \zeta$ dB) is written above the bars in (b,c).
    Initial occupation of the mechanical oscillator $n_0 = 0.05$.
    Gray dashed lines show Wigner functions of ideal Fock states $\ket{1}$ and $\ket{2}$.
    Numbers near the dips of the Wigner functions show the loss $\zeta$ (in dB) required to lose the negativity.
    (b,c) multiphoton probabilities of the state of light at the detector corresponding to lines in (a).
    Dashed horizontal lines show thresholds of multiphoton QNG.
  }
  \label{fig:non-gauss-readout}
\end{figure}

The quantum state of the optimal detector mode is studied at~\cref{fig:non-gauss-readout}.
First, the estimations show that with optimal set of parameters, the optimal detector mode is, in principle, capable of almost a perfect state swap from the mechanics ($\cT \approx 1$).
Consequently, the limiting factor for the verification is the nonzero loss $\zeta$.
In~\cref{fig:non-gauss-readout}, we visualize the detector mode states showing their Wigner functions and multiphoton probabilities.
We can see that the determinative factor of the possibility to verify the QNG character of the mechanical state is the efficiency of the detection scheme $1 - \zeta$.
Curiously, due to the presence of higher Fock states contributions in the conditional states, slightly more than $\SI{50}{\percent}$ of loss rids the conditional Wigner functions of negativity
(an attenuated single-photon state $\zeta \dyad{0}{0} + ( 1 - \zeta ) \dyad{1}{1}$ loses its negativity exactly at $\zeta = 1/2 \approx \SI{3.01}{\dB}$).

\subsection{Estimation of phase-randomized displacement} \label{sec:estimation_of_phase_randomized_displacement}

\begin{figure}
  \begin{center}
    \includegraphics[width=0.99\linewidth]{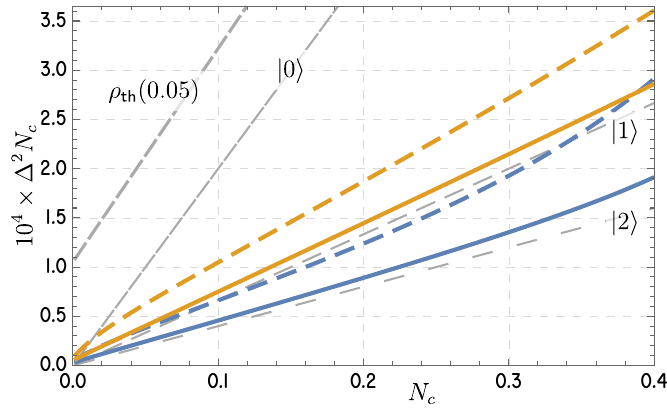}
    \renewcommand{\arraystretch}{1.3}
    \begin{tabular}{@{}lcccccc@{}}
      \toprule
      State                                       & $\ket{m}$                  & $\ket{0}$ & $\CA^2 [\ket{0}] $ & $\CA^2 [\hat \rho_*] $ & $\CA[ \ket{0} ]$ & $\CA [\hat \rho_*]$ \\
      \midrule
      $10^4 \times \dfrac{ \Delta^2 N_c }{ N_c }$ & $\dfrac{ 20 }{ ( 2m + 1)}$ & $20$      & \num{4.58}         & \num{6.68}           & \num{7.00}       & \num{8.48}        \\
      \bottomrule
    \end{tabular}
  \end{center}
  \caption{Estimation error of phase-randomized displacement assuming $M=500$ copies of the probe state.
  Color lines correspond to phonon-added thermal states, solid lines: to the initial ground state, dashed lines: to initial thermal state with mean occupation $n_0 = 0.05$ phonons.
  Yellow lines: single phonon added, blue lines: two phonons added.
  Gray lines of various dashing correspond to the example probe states, as indicated by the labels: Fock states and thermal state $\hat \rho_* = \rho\s{th}(0.05)$ with mean occupation $0.05$.
  Inset shows the magnified area of low $N_c$.
  Table below indicates the coefficient of linear fit of the estimation error as a function of $N_c$ in the region $N_c \leqslant 0.3$ where the dependencies are approximately linear.
  }
  \label{fig:fisher_information}
\end{figure}

In this section, we consider theoretically sensing a phase-randomized displacement~\cite{oh_optical_2020} using the phonon-engineered states as an example of their practical applicability.
We show that adding one or two phonons to a thermal state significantly increases its sensing capabilities.
For full derivations, see~\cref{sec:detection_of_phase_randomized_displacement}.

A phase-randomized displacement of magnitude $\sqrt{ N_c}$  changes the probe state $\rho\s{in}$ to $\rho\s{f}$ that reads
\begin{multline}
  \rho\s{f} = \int_0^{ 2 \pi } \frac{ \dl \phi }{ 2 \pi }
  \hat D ( \sqrt{ N_c } \ee^{ \ii \phi } ) \rho\s{in} \hat D\dg (\sqrt{ N_c } \ee^{ \ii \phi }) =
  \\
  \sum_n p\s f (n | N_c ) \dyad{ n }{ n }.
\end{multline}
The last equation represents the output state $\rho\s{f}$ is the basis corresponding to the phonon-number resolved detection characterized by the POVM $\dyad{k}{k},\ k \in \mathbb N_0$.
Accordingly, the Fisher information of the probe state reads~\cite{oh_optical_2020}
\begin{equation}
  \label{eq:displ:fisher}
  F(N_c) = \sum_n \frac{ 1 }{ p\s{f} (n | N_c ) } \mleft( \frac{ \partial p\s{f} (n| N_c) }{ \partial N_c } \mright)^2.
\end{equation}
The minimal mean-sqare error $\Delta^2 N_c$ of estimation of $N_c$ attainable by any unbiased estimator is then given by the quantum Cramér-Rao inequality~\cite{helstrom_quantum_1969,braunstein_statistical_1994}
\begin{equation}
  \Delta^2 N_c \geq \frac{ 1 }{ M F(N_c) },
\end{equation}
where $M$ is the number of copies of the probe state used in the estimation procedure.

Our simulations of the heralded states' capabilities are in~\cref{fig:fisher_information}.
On average, the more phonons are added to the initial thermal state, the lower is the estimation error.
The purity of the initial state also increases the sensitivity of the heralded state.
Both single- and two-phonon added thermal states significantly outperform the original thermal state.
The error is approximately linear in $N_c$ for small values of $N_c$ (equivalently, relative error does not depend on $N_c$).
The coefficients of the linear dependence for different probe states are also presented in a table in the caption of~\cref{fig:fisher_information}.
Importantly, perfect phonon-number resolution is not required to achieve this result, the observed levels of errors are reachable already with resolution of up to 2 phonons for single-phonon-added states, and up to 3 phonons for two-phonon-added states.
For details, see~SM.

\section{Discussion and Conclusion} \label{sec:discussion_and_conclusion}

In this manuscript, we propose a scenario for an in-detail study of the quantum non-Gaussian nature of multiphonon-added quantum states of mechanical oscillations.
Drawing inspiration from the system in~\cite{patil_measuring_2022a} that has performed heralding additions and subtractions of phonons to the motion of superfluid Helium, we prove that with feasible parameters, conclusive observation of QNG is possible.
Cooling below a feasible mean occupation of $n_0 = 0.2$ allows generation of mechanical states with one- or two-phonon QNG via addition of a single or two phonons.
Given an available cooling down to a lower mean occupation of approximately $n_0 = 0.05$, it is possible to observe the two-phonon QNG in the leaking light.

Addition of one or two phonons to the thermal state with mean occupation \num{0.05} significantly enhances its performance for phase-randomized displacement sensing.
Further cooling down prior the additions, ultimately to the ground state of mechanical oscillator, brings larger advantage for such sensing.
Importantly, resolution of only a few photons in the detector is needed (up to three for two-phonon-added states, or up to two for single-phonon-added states).

The main limiting factor in generation and observation of QNG is the thermalization to the mechanical environment.
Suppressing the decoherence further would allow continued longer sequences of heralding events reaching higher Fock states, and would improve the purity of conditional mechanical states.

The analysis presented here stimulates further optomechanical developments in experimental investigations of essential quantum non-Gaussian coherence in the superposition of Fock states needed for quantum sensing of phase~\cite{mccormick_quantumenhanced_2019} and binomial error correction codes~\cite{hu_quantum_2019}, as already achieved for the trapped ions and superconducting circuits.
Moreover, multimode mechanical quantum non-Gaussian coherences can be used to advance towards Duan-Lukin-Cirac-Zoller types of communication protocols~\cite{riedinger_remote_2018,zivari_onchip_2022} and mechanical Bell tests~\cite{marinkovic_optomechanical_2018} in a new regime, and more efficient dual and multi-rail error correction codes~\cite{koottandavida_erasure_2024} developed recently at different platforms.
Optical control and efficient readout of such unique mechanical systems in a quantum non-Gaussian regime will allow us to overcome the limitations of trapped ions and superconducting experiments in their interconnections.

\section{Methods} \label{sec:methods}

\subsection{Description of the open optomechanical system dynamics} \label{sec:description_of_dynamics_of_open_optomechanical_system}

The standard approach to weakly coupled optomechanical cavities is by using Heisenberg-Langevin equations of motion~\cite{genes_quantum_2009,aspelmeyer_cavity_2014,khalili_quantum_2016}.
The Hamiltonian of the system reads (assuming the interaction is linearized around strong mean classical amplitudes and $\hbar = 1$)
\begin{equation}
  H = \Delta a\s{c}\dg a\s{c} + \omega_m a\s{m} \dg a\s{m} - g ( a\s{c}\dg + a\s{c})( a\s{m}\dg + a\s{m} ),
  \label{eq:methods:hamiltonian}
\end{equation}
where $a\s{c,m}$ are the annihilation operators of the cavity mode and the mechanical motion, $\Delta = \omega\s{c} - \omega\s{d}$ is the detuning between the drive and cavity frequency, $g = \omega\s{c}' (x_{m} )  \sqrt{\ev{ n\s{c} }}$ is the optomechanical coupling rate enhanced by the mean photon number $\ev{n\s{c}}$ due to the classical drive, $\xmm = a\s{m} + a\s{m}\dg$ is the dimensionless displacement of the mechanical oscillator.
In presence of strong classical drive at (anti-) Stokes sideband, $\Delta = \pm \omega\s{m}$, the Hamiltonian is further simplified (in the frame defined by the first two terms of~\eqref{eq:methods:hamiltonian}) under rotating wave approximation:
\begin{equation}
  \label{eq:methods:sideband:ham}
  H_{\Delta = \omega\s{m}} = g a\s{c}\dg a\s{m} + \hc,
  \quad
  H_{\Delta = - \omega\s{m}} = g a\s{c}\dg a\s{m}\dg + \hc
\end{equation}

Using~\cref{eq:methods:sideband:ham} one can write the equations of motion, e.g. for $a\s{c,m}$:
\begin{subequations}
\begin{align}
  \dot a\s{c} &= \ii \comm{ H }{ a\s{c} } - \kappa a\s{c} + \sqrt{ 2 \kappa } a\s{in},
  \\
  \dot a\s{m} &= \ii \comm{ H }{ a\s{m} } - \frac{ \gamma}{2 } a\s{m} + \sqrt{ \gamma } a\s{th}.
\end{align}
\label{eq:methods:HLE}
\end{subequations}
Here $\kappa$ is the cavity linewidth, $\gamma$ is the mechaincal viscous damping rate, $a\s{in,th}$ are the input quantum noises.

Solution of~\eqref{eq:methods:HLE} gives the instantaneous values of $a\s{c,m}$.
To describe the field leaking from the cavity towards the detector, we use input-output relations~\cite{collett_squeezing_1984}
\begin{equation}
  a\s{out} (t) = - a\s{in} (t) + \sqrt{ 2 \kappa } a\s{c} (t).
\end{equation}

Finally, the detector mode is described by the ladder operators that can be derived from $a\s{out}$ using~\cref{eq:verification:modedef}.

\subsection{Gaussian statistics and operations} \label{sec:gaussian_statistics_and_operations}

\Cref{sec:description_of_dynamics_of_open_optomechanical_system} outlines the procedure to obtain input-output relations that connect operators describing the initial state of the optomechanical system and quantum inputs, and the output state of the system.
Since the input states are all Gaussian and the dynamics are linear, prior the detection, the state of the system remains Gaussian, and can be described using its first two moments.

In particular, the initial state of the cavity and the state of input optical noise is vacuum, the initial state of the mechanical oscillator, and the input mechanical noise are both in thermal states, with mean occupations, respectively, $n_0$ and $n\s{th}$.

The output state of the optomechanical interaction prior the detection at time $\tau$ is a Gaussian state of the bipartite system, formed by the mechanical oscillator with quadratures $x\s{m}$ and $p\s{m} = ( a - a\dg )/ \ii$, and the detector mode with quadratures $\Mode X\s{L} = \Mode A\s{det} + \Mode A\s{det}\dg$ and $\Mode P\s{L} = (\Mode A\s{det} - \Mode A\s{det}\dg)/ \ii$.
Covariance matrix of this system corresponding to the vector of quadratures $\mleft(x\s{m}, p\s{m}, \Mode X\s{L}, \Mode P\s{L}\mright)$ can be written in the block form as
\begin{equation}
  \mmat V =
  \begin{pmatrix}
    \mmat C\s{m,m} & \mmat C\s{m,L}
    \\
    \mmat C\s{m,L}\trp & \mmat C\s{L,L}
  \end{pmatrix}.
\end{equation}
Mean values of the quadratures are all zeroes, thanks to the zero-mean initial states and input noises.

The detector mode is then registered by an on-off detector, which is characterized by a POVM with two elements: $\Pi_0 = \dyad 00$ and $\Pi_1 = \eye - \Pi_0$.
The operator $\Pi_0$ projects the detector mode on vacuum which is a Gaussian operation.
The action of $\Pi_1$ is not Gaussian, however, it can be written as a negative combination of two Gaussian operations (see details in SM).

The conditional state of the mechanical oscillator, heralded by a click of the detector, can be written as follows, denoting by $\rho\s{G} ( \mmat V )$ density matrix of a Gaussian state with covariance matrix $\mmat V$.
\begin{equation}
  \label{eq:methods:conditstate}
  \mleftright
  \rho\s{cond} = \frac{ 1 }{ 1 - p\s{off} }
  \left[ \rho\s{G} (\mmat C\s{m,m}) - p\s{off} \rho\s{G} (\mmat C\s{m,m}' )\right],
\end{equation}
where $p\s{off} = 2 \mleft[ \det( \mmat C\s{det} + \eye_2 )\mright]^{-1/2}$ is the probability of the detector not producing a click (``off'' result).
The conditional covariance matrix $\mmat C\s{m}'$, corresponding to the detection of vacuum in the leaking light, reads
\begin{equation}
  \mmat C\s{m,m}' = \mmat C\s{m,m}  - \mmat C\s{m,L} ( \mmat C\s{L,L} + \eye_2 ) ^{-1} \mmat C\s{m,L}\trp.
\end{equation}

For a generalization of this detection scheme to include imperfect detection, a Hanbury-Brown and Twiss measurement, and for description of multiple subsequent heraldings, see SM.

\subsection{Evaluation of non-Gaussianity via multiphonon contributions} \label{sec:evaluation_of_non_gaussianity}

The conditional state~\eqref{eq:methods:conditstate} is a [negative] mixture of two thermal states.
It is possible to derive effective occupation numbers $n\s{m}, n\s{m}'$ such that
\begin{equation}
  \label{eq:methods:condoccup}
  \mmat C\s{m,m} = ( 2 n\s{m} + 1 ) \eye_2,
  \quad
  \mmat C\s{m,m}' = ( 2 n\s{m}' + 1 ) \eye_2.
\end{equation}
In Fock basis, the diagonal elements of thermal states' density matrices
\begin{equation}
  \label{eq:methods:pktherm}
  p_k = \mel{k}{ \rho\s{G}(  ( 2 n + 1 )\eye_2 ) }{ k } =
  \frac{ n^k }{ ( n + 1 )^{k + 1} }.
\end{equation}
Combining~\cref{eq:methods:conditstate,eq:methods:condoccup,eq:methods:pktherm}, we can write
\begin{multline}
  \mel{k}{\rho\s{cond}}{k}
  \\
  =
  \frac{ 1 }{ 1 - p\s{off}}
  \mleft[
    \frac{ n\s{m}^k }{ ( 1 + n\s{m} )^{k + 1 }} -
    p\s{off} \frac{ ( n\s{m}' )^k }{ ( 1 + n\s{m}' )^{k + 1 }}
  \mright].
\end{multline}

\subsection{Parameters for the simulation} \label{sec:parameters_for_simulation}

The parameters we use for simulation are extracted from~\cite{patil_measuring_2022a,wang_manipulating_2023} and summarized in~\cref{tab:params}.

\begin{table}[htb]
\begin{tabular}{@{}lll@{}} \toprule
  Parameter                   &                     & Value                                   \\ \midrule
  Mechanical frequency        & \(\omega\s{m}\)     & \( \angfreq{315.3}{\mega} \)            \\
  Mechanical damping          & \(\gamma\)          & \( \angfreq{3.12}{\kilo} \)             \\
  Cavity linewidth            & \(2 \kappa\)        & \( \angfreq{47.2}{\mega} \)             \\
  Single-photon coupling      & \(g_0\)             & \( \angfreq{4.6}{\kilo} \)              \\
  Intracavity power           & \( \mathcal P_0 \)  & \( \SI{1}{\micro\watt} \)               \\
  Intracavity photon number   & \( n\s{cav} \)      & \( \num{1e6} \)                         \\
  Enhanced coupling           & \(g \)              & \( \leq \angfreq{1}{\mega} \)           \\
  Blue-detuned pulse duration & \(\tau\s b\)        & \( \SIrange{2}{3.5}{\micro\second} \)   \\
  Red-detuned pulse duration  & \(\tau\s r\)        & \( \SIrange{5.5}{9.5}{\micro\second} \) \\
  Blue-detuned pulse duration & \(\kappa \tau\s b\) & \numrange{6e2}{1.1e3}                   \\
  Red-detuned pulse duration  & \(\kappa \tau\s r\) & \numrange{1.6e3}{2.8e3}                 \\
  Environment temperature     & \(T\s{bath} \)      & \( \SI{20}{\milli\kelvin} \)            \\
  Environment occupation      & \(n\s{bath} \)      & \( \num{1.3}    \)                      \\
 \bottomrule
\end{tabular}
\caption{Numerical parameters of the superfluid experiment used for the simulations}
\label{tab:params}
\end{table}

\section{Acknowledgement} \label{sec:acknowledgement}

The authors thank Jack Harris and Yogesh Patil for helpful discussions.
We acknowledge the project 23-06308S of the Czech Science Foundation and project CZ.02.01.01/00/22\_008/0004649 of the MEYS Czech Republic supported by the EU funding.
R.F. also acknolwedges funding from the MEYS of the Czech Republic (Grant Agreement 8C22001), Project SPARQL has received funding from the European Union's Horizon 2020 Research and Innovation Programme under Grant Agreement no. 731473 and 101017733 (QuantERA).

\cleardoublepage
\appendix

\section{Addition and subtraction of single excitations by on-off detectors} \label{sec:addition_and_subtraction_of_single_excitations_by_on_off_detectors}

The technique to conditionally add or subtract a single excitation by optical heralding is based on the following principles.

Consider a mechanical mode, initially in a state $\ket{\psi\s{in}}$, interacting with an optical mode, initially in vacuum, via the Hamiltonian of the form (\cref{fig:fig0}~b)
\begin{equation}
  \label{eq:appcond:hamiltonian}
  H\s{om} = g (  \xi\s{m}  a\s{c}\dg +  \xi\s{m}\dg a\s{c} ),
\end{equation}
where $g$ is the coupling rate, and $\hat \xi\s{m}$ is a ladder operator of the mechanical oscillator.
In linearized optomechanics, driving on the upper mechanical sideband yields the Hamiltonian~\cref{eq:appcond:hamiltonian} with $\xi\s{m} = a\s{m}\dg$, and driving on the lower sideband, $\xi\s{m} = a\s{m}$ (see~\cref{fig:fig0}~b,c).
Interaction for an interval $\tau$ under the action of this Hamiltonian will transforms the states of the two modes by the action of the unitary operator
\begin{multline}
  U\s{om} = \ee^{ - \ii g \tau (  \xi\s{m}  a\s{c}\dg +  \xi\s{m}\dg a\s{c} ) }
  \\
  \approx
  \eye - \ii g \tau (  \xi\s{m}  a\s{c}\dg +  \xi\s{m}\dg a\s{c} ) + o (g \tau)^2,
\end{multline}
where we assume a weak interaction over a short time.
The bipartite state then is transformed as
\begin{equation}
  \ket{ 0 \s{c} } \otimes \ket{ \psi\s{in} } \mapsto \ket{ 0 \s{c} } \otimes \ket{ \psi\s{in} }
  - \ii g \tau  \ket{1\s{c}} \otimes (  \xi\s{m}  \ket{\psi\s{in}} ).
\end{equation}
A consequent detection by the on-off detector (the POVM element $\qeye - \dyad 00$) removes the vacuum contribution (the first term), and the conditional mechanical state (up to normalization) reads
\begin{equation}
  \xi\s{m} \ket{ \psi\s{in} }.
\end{equation}

The assumptions necessary for this derivation include: (i) weak coupling and short interaction time: $g \tau \ll 1$, (ii) unitary optomechanical interaction, which is equivalent to low mechanical noise $\gamma \tau n\s{th} \ll 1$ and $\gamma \ll g$, (iii) good sideband resolution $\omega\s{m} \gg \kappa$ necessary to resolve Stokes and anti-Stokes photons.
We also implicitly eliminated the cavity mode, assuming that $a\s{c}$ is the operator that describes the detector mode, which requires pulse durations sufficiently long for the scattered by optomechanical interaction photons to leak from the cavity: $\kappa \tau \gg 1$.

\section{Derivation of conditional quantum states of mechanical oscillator} \label{sec:heisenberg_langevin_dynamics}

\subsection{Equations of motion} \label{sec:equations_of_motion}

Following the standard approach in optomechanics theory~\cite{aspelmeyer_cavity_2014}, we use the linearized Heisenberg-Langevin equations of motion to study the dynamics of the optomechanical system.
In the Heisenberg picture, a quantum state of the system is characterized by a vector of quadratures
\( \mvec r = ( \xmm, \pmm , \xll, \pll ) \) with commutation relations $\comm{ \hat X_i }{ \hat P_i } = 2\ii$.
The equation of motion for this vector reads
\begin{equation}
  \label{eq:heisenberg:langevin}
  \diff{\mvec r }{t } = \mmat A \mvec r + \mvec r\up{N},
\end{equation}
where $\mmat A$ is the drift matrix, $r\up{N}$ is the vector of quantum noise inputs, including thermal noise of the mechanical environment and the optical input noise.
In the resolved sideband regime (\(\omega\s m \gg \kappa\)), assuming driving on a mechanical sideband (\( \Delta = \pm \omega \s m\)), and a weak coupling (\(g \leqslant \kappa\)), the expressions for the drift matrix reads
\begin{equation}
  \mmat A =
  \begin{pmatrix}
  -\frac{\gamma}{2} & 0                 & 0             & g\s B -g\s R \\
  0                 & -\frac{\gamma}{2} & g\s B + g\s R & 0            \\
  0                 & g\s B -g\s R      & -\kappa       & 0            \\
  g\s B + g\s R     & 0                 & 0             & -\kappa      \\
\end{pmatrix}.
\end{equation}
Here $\gamma$ is the viscous damping rate of the mechanical mode and $\kappa$ is the optical linewidth.
We also formally define the effective linearized coupling rates while driving on the blue or red sideband $g\s{B,R} = g_0  \ev{ n\up{cav}\s{B,R}}^{1/2}$, where $\ev{n_\bullet \up{cav}}$ is the mean photon number in the cavity due to the drive.
Note that these quantities are not simultaneously nonzero: when driving on the lower sideband, $g\s{B} = 0$ and when driving on the upper sideband, $g\s R =  0$.
The noise vector has elements
\begin{equation}
  \mvec r\up{N} = \mleft(
  \sqrt{ \gamma} x\up{th},
  \sqrt{ \gamma} p\up{th},
  \sqrt{ 2 \kappa} x\up{in} ,
  \sqrt{ 2 \kappa} p\up{in}
  \mright),
\end{equation}
with the first two elements being quadratures of the thermal noise and the last two --- of the input optical fluctuations.
These quadratures satisfy commutation relations
\begin{equation}
  \comm{ x^i (t) }{ p^i (t') } = 2 \ii \delta ( t - t' ),
  \qquad
  i = \scrpt{in,th}.
\end{equation}

\cref{eq:heisenberg:langevin} allows a formal solution of the form
\begin{equation}
  \label{eq:heislang:gen:sol}
  \mvec r(t) = \mmat M (t) \mvec r (0) + \int_0^t \dl{s} \mmat M(t - s) \mvec r\up{in} (s),
\end{equation}
where \(\mmat M (t) \equiv \exp[ \mmat A t ]\).
The solution readily provides an expression for the instantaneous values of the quadratures of the mechanical oscillator as a function of time.
Using input-output relations for the light~\cite{collett_squeezing_1984}
\begin{equation}
  \label{eq:io:ladder}
  a\up{out}(t) = - a\up{in}(t) + \sqrt{ 2 \kappa } a\up{cav}(t),
  \text{ where }
  a^i = \frac 12 ( x^i + \ii p^i ),
\end{equation}
one can have solutions for the instantaneous amplitude of the leaking light $a\up{out}(t)$.
In order to deal with bosonic statistics of the output light, we have to specify the temporal modes of interest of the output light~\cite{christiansen_interactions_2023}.
This is done by introducing the temporal filtering
\begin{equation}
  \label{eq:methods:defdetmode}
  \Mode A\up{det} (t) = \int_{t - \tau}^t \dl{s} a\up{out} (s) f\up{out} (s),
\end{equation}
where $f\up{out}$ is the \emph{filtering function}.
A proper normalization of $f\up{out}$ ensures the canonical commutation relations for the filtered temporal mode:
\begin{equation}
  \int_{t - \tau}^t \dl{s} \abs{ f\up{out} (s) }^2 = 1
  \Rightarrow
  \comm{\Mode A\up{det} (t) }{ (\Mode A\up{det}(t) )\dg } = 1.
\end{equation}
Correspondingly, the canonical quadratures of this mode, $(\Mode X\s{L}, \Mode P\s{L}) = (\Mode A\up{det} + (\Mode A\up{det} )\dg, \ii ( ( \Mode A\up{det} )\dg - \Mode A\up{det} ) )$ satisfy the standard commutation relation $\comm{ \Mode X\s{L} }{ \Mode P\s{L}} = 2 \ii$.

One natural choice of the filtering function is a flat-top pulse: $f\up{out} = \tau^{-1/2}$.
The temporal modes corresponding to this filtering function are detected by, e.g., by a homodyne detector with a local oscillator with time-independent amplitude.
A more optimal choice~\cite{rakhubovsky_nonclassical_2019}, from the point of view of reconstruction of the state of the mechanical oscillator, can be done by deriving $f\up{out}$ based on the formal solution~\eqref{eq:heislang:gen:sol} of the equations of motion.
By substituting~\cref{eq:heislang:gen:sol} into~\cref{eq:io:ladder}, one gets that the quadratures of the mechanical oscillator enter the expression for the output light as
\begin{equation}
  a\up{out} (t) + ( a\up{out} (t) )\dg \propto \cdots + \mmat M_{3,1} (t) X\s{m} (0).
\end{equation}
Accordingly, choosing (up to normalization)
\begin{equation}
  f\up{out}(t) = f\up{opt}(t) \propto \mmat M_{3,1} (t),
\end{equation}
maximizes the contribution of $X\s{m} (0)$ in $\Mode X\s{L}(t)$.
In practice, such filtration can be approximated by postprocessing of the results of frequent sampling of the detection with flat-top filtering function.
\begin{equation}
  \Mode X\s{L} = \lim_{\Delta t \to 0} \sum_i
  \frac{ C_i }{ \sqrt{ \Delta t }} \int_{t_i - \Delta t}^{t_i} \dl {s} X\up{out} (s),
\end{equation}
where $C_i$ are chosen appropriately (e.g. $C_i \approx f\up{opt} (t_i)$).

\subsection{Effective Input-Output relations for second moments} \label{sec:effective_input_output_relations}

Solution of the Heisenberg-Langevin equations allows to effectively describe the dynamics of the optomechanical system by the relations that map quantum inputs to the outputs following the spirit of~\cite{gardiner_quantum_2004}.
In practice, the solution for the state of the optomechanical system at the end of a pulse (below for brevity of notation we write it at time $t = \tau$ assuming the pulse begins at $t = 0$) is determined by the initial state of the intracavity field and the mechanical oscillator, and by the quantum state of the input noises.
It is convenient to write the input-output relations for the mechanics and the output light separately.
We first introduce notation
\begin{multline}
  \mvec r\up{out} (\tau)
  =
  \mleft( X\s{m} (\tau), P\s{m} (\tau), \Mode X\s{L} (\tau) , \Mode P\s{L} (\tau) \mright)
  \\
  =
  \mvec r\s{m} (\tau) \oplus \mvec r\s{L} (\tau),
\end{multline}
where $\mvec r_k$ is a vector containing quadratures of the $k$-th mode, e.g. $\mvec r\s{m} (0) = (X\s{m} (0), P\s{m} (0))$.
Then, the solution of the Heisenberg-Langevin equation reads
\begin{multline}
  \mvec r_k
  = \sum_i u\s{k,i} (\tau)\mmat R (\theta_{k,i } ) \mvec r_i ( 0 )
  \\ +
  \sum_j \int_0 ^\tau \dl{s}  v\s{k,j} ( s )\mmat R ( \theta_{k,j } ) \mvec r_j (s),
  \label{eq:io:quads}
\end{multline}
where $k$ can take values $\scrpt{m,L}$, $i$ can take values $\scrpt{m, c}$, and $j$ can take values $\scrpt{in, th}$.
The scalar functions $u_i$ and $v_j$ are determined by the parameters of the optomechanical interaction, and by the choice of the filtering function.
The full form of these functions is too long to present them here, however, they can be straightforwardly derived from~\cref{eq:heislang:gen:sol,eq:io:ladder,eq:methods:defdetmode}.
E.g.~$u\s{m,c} (\tau) = \mmat M_{1,3} (\tau)$.
Finally, rotation matrices $\mmat R(\theta)$ indicate a possible phase rotation introduced by the optomechanical interaction.

The transformations described by~\cref{eq:io:quads} are linear in quadrature operators which is a consequence of quadratic Hamiltonian of the system.
Given that the initial state of the system is Gaussian, including Gaussian initial states of both intracavity light and mechanical oscillator, as well as Gaussian states of input noises, the output state of the system is also Gaussian.
Gaussian states are such states whose Wigner function~\cite{schleich_quantum_2001} is a Gaussian function.
These states can be fully described by the first two moments of the quadratures.
In our special case, the output state has zero mean values of all the output quadratures, and therefore, can be fully described by its covariance matrix, the matrix with elements
\begin{equation}
  \label{eq:methods:covmatoutput}
  \mmat V\s{out} = \cov (\mvec r\up{out}, \mvec r\up{out})
  =
  \begin{pmatrix}
    \mmat C\s{ m, m } & \mmat C\s{ m, L } \\
    \mmat C\s{ m, L }^\intercal & \mmat C\s{ L, L }
  \end{pmatrix}.
\end{equation}
where we define the operation of taking covariance as
\begin{equation}
\left[ \cov (\mvec x, \mvec y) \right]_{ij } \colonequals
  \frac 12 \ev{ \mvec x_i \mvec y_j + \mvec y_j \mvec x_i } - \ev{ \mvec x_i}\ev{ \mvec y_j },
\end{equation}
and
\begin{equation}
  \mmat C\s{A, B} = \cov( (X\s A, P\s A) , (X\s B, P\s B)),
  \qquad
  \scrpt{A,B} = \scrpt{m,L}.
\end{equation}

In order to compute the covariance matrix from~\cref{eq:io:quads}, one must specify the statistics of all $\mvec r_i$ entering it.
In our computations, we assume the intracavity light to be initially in vacuum, the initial mechanical state to be a thermal state with mean occupation $n_0$, the input optical and mechanical noise to be Markovian, respectively, in the ground state and in a thermal state with mean occupation $n\s{th}$.
This can be summarized as
\begin{subequations}
\begin{align}
  \cov( \mvec r\s{c} (0) , \mvec r\s{c} (0) )    & = \sigma\s{c} \eye_2,
  \\
  \cov( \mvec r\s{m} (0) , \mvec r\s{m} (0) )    & = \sigma\s{m} \eye_2,
  \\
  \cov( \mvec r\s{in} (t),  \mvec r\s{in} (t') ) & = \sigma\s{in} \eye_2 \delta (t - t'),
  \\
  \cov( \mvec r\s{th} (t),  \mvec r\s{th} (t') ) & = \sigma\s{th} \eye_2 \delta(t - t'),
\end{align}
\label{eq:methods:initcovms}
\end{subequations}
where
\begin{gather}
  \sigma\s{c} = \sigma\s{in} = 1,
  \\
  \sigma\s{m} = 2 n_0 + 1,
  \quad
  \sigma\s{th} = 2 n\s{th} + 1,
\end{gather}
and the averaging is over the initial state and the state of the environment.

The expressions for $\mmat C_{i, j}$ then read
\begin{equation}
  \mmat C_{i,i } = \sum_j  T_{i,j } \sigma_j \eye_2,
  \quad
  \mmat C\s{m,L} = \sum_j c_j \sigma_j
  \begin{pmatrix}
    0 & 1 \\ - 1 & 0
  \end{pmatrix},
\end{equation}
where $i$ takes values $\scrpt{m, L}$ and $j$ takes values $\scrpt{c, m, in, th}$.
The coefficients $T_{i,j }$ and $c_j$ indicate the contributions of the quantum inputs of the optomechanical system in its outputs.
These coefficients are functions of the pulse duration, the configuration of the optomechanical setup (coupling rate, detuning etc.) and the filtering function that defines the output mode of interest.

\subsection{Heralding using on-off detectors} \label{sec:heralding_using_on_off_detectors}

Evolution governed by the optomechanical interaction maps initial Gaussian states onto different Gaussian states.
The non-Gaussianity in our system emerges from detection of light using highly nonlinear detectors.
Avalanche photodetectors (APDs) perform inherently non-Gaussian measurements, and can be characterized by the POVM consisting of two Hermitian operators
\begin{equation}
  \hat \Pi_0 = \dyad{0\s{L}}{0\s{L}},
  \qquad
  \hat \Pi_1 = \hat \II\s{L} - \hat \Pi_0,
\end{equation}
where $\hat \II\s{L}$ is an identity operator in the Hilbert space corresponding to the light mode at the detector.
Since $\hat \Pi_0$ corresponds to a Gaussian detection (detecting vacuum), the action of this detector can be described using tools of Gaussian quantum information.

Let us apply this knowledge to describe the optomechanical heralding.
The conditional mechanical state after the detection event reads
\begin{equation}
  \rho\s{m}\up{cond} =
  \frac{ \Tr\s{L} [ \rho\s{mL} \Pi_1  ]}{ \Tr\s{m,L} [ \rho\s{mL} \Pi_1 ]} =
  \frac{ \Tr\s{L} [ \rho\s{mL} \eye\up{m} ( \eye\up{L} - \Pi_0 )  ]}{ \Tr\s{m,L} [ \rho\s{mL}  \eye\up{m} ( \eye\up{L} - \Pi_0 ) ]}.
\end{equation}

To further simplify this expression, consider a multimode Gaussian state with covariance matrix in block form
\begin{equation}
  \mmat V\s{pre} =
  \begin{pmatrix}
    \mmC\s{A,A}           & \mmC \s{A,B} \\
    \mmC\s{A,B}^\intercal & \mmC \s{B,B}
  \end{pmatrix},
\end{equation}
where A and B are collections of modes such that there are $N$ modes in B.
A measurement on the modes B that projects these modes on another Gaussian state with CM $\mmat R\s{B,B}$, heralds the modes A being in a \emph{conditional} state
\begin{equation}
  \rho\s{cond} (\mmat V\s{pre}, \mmat R\s{B,B}) = \rG ( \mmat V\s{cond} (\mmat V\s{pre}, \mmat R\s{B,B})).
\end{equation}
This is a Gaussian state with CM
\begin{equation}
  \mmat V\s{cond}\up{A} (\mmat V\s{pre}, \mmat R\s{B,B}) = \mmC\s{A,A} - \mmC \s{A,B} ( \mmC \s{B,B} + \mmat R\s{B,B} )^{-1} \mmC \s{A,B}^\intercal.
\end{equation}
The probability of this detection is given by
\begin{equation}
  p\s{cond}\up{B}( \mmat V\s{pre}, \mmat R\s{B,B} ) = \frac{ 2^N }{ \sqrt{ \det( \mmC\s{B,B} + \mmat R\s{B,B} ) }}.
\end{equation}

From this follows in particular that (skipping $\eye\up{m}$ for brevity)
\begin{gather}
  \Tr\s{m,L} \rho\s{mL} \Pi_0 = p\s{cond}\up{L} ( \mmat V, \eye_2 ) = p\s{off},
  \\
  \Tr\s{L} \rho\s{mL} \Pi_0 = p\s{off} \rG ( \mmat V\s{cond}\up{m} ( \mmat V , \eye_2 ) ) = p\s{off} \rG ( \mmat C'\s{m,m}),
  \\
  \mmat C\s{m,m}' = \mmat C\s{m,m} - \mmat C\s{m,L} (\mmat C\s{L,L} + \II_2 )^{-1} \mmat C\s{m,L}^\intercal,
  \\
  \Tr\s{L} \rho\s{mL} \eye\up{L} = \rG ( \mmat C\s{m,m} ),
  \\
  \Tr\s{m,L} \rho\s{mL} \eye\up{m} \eye\up{L} = 1.
\end{gather}

Finally, the conditional mechanical state reads
\begin{equation}
  \label{eq:apd:rho:cond}
  \rho\up{cond}\s{m} = \frac{1}{ 1 - p\s{off} } [ \rho\s{G} (\mmat C\s{m,m} ) -  p\s{off} \rho\s{G} (\mmat C'\s{m,m})].
\end{equation}

For a generalization to an Hanbury-Brown and Twiss scheme allowing two simultaneous additions/subtractions, see~\cref{sec:instantaneous_multiple_additions_subtractions}.
 
\subsection{Evolution of heralded non-Gaussian states} \label{sec:evolution_of_heralded_non_gaussian_states}

The state~\eqref{eq:apd:rho:cond} is a negative mixture of two Gaussian (thermal) states.
Importantly, the result of action of a quantum map $\Lambda[ \bullet ]$ on such a sum can be distributed over the individual terms:
\begin{equation}
  \mleftright
  \Lambda\left[ \sum_i p_i \rG (\mmat V_i ) \right]
  =
  \sum_i p_i \Lambda[ \rG (\mmat V_i ) ].
\end{equation}

For instance, a single-mode Gaussian map that does not cause phase-space displacement, can be written as transforming the covariance matrix as
\begin{equation}
  \label{eq:methods:covmatIO}
  \mmat V \mapsto \mmat T \mmat V \mmat T\trp + \mmat N.
\end{equation}
The matrices $\mmat T$ (effective transmittance) and $\mmat N$ (added noise) fully describe the map.
In particular, for the case of decoherence due to coupling at rate $\gamma$ over an interval with duration $\tau\s{del}$ to a thermal bath with mean occupation $n\s{th}$, these matrices read
\begin{equation}
  \label{eq:methods:decoherence}
  \mmat T\s{th} = \ee^{ - \gamma \tau\s{del} } \eye_2,
  \quad
  \mmat N\s{th} = ( 1 - \ee^{ - \gamma \tau\s{del}} ) ( 2 n\s{th} + 1 ) \eye_2.
\end{equation}

Furthermore, the map $\Lambda[ \bullet ]$ needs not to be Gaussian, it can describe a non-Gaussian operation such as a phonon addition/subtraction.
To ensure that there are no ``stray'' optomechanical correlations with the intracavity field which is implicitly eliminated from the covariance matrix~\eqref{eq:methods:covmatoutput}, we assume sufficient delay between subsequent pulses.
Indeed, the intracavity field leaks and gets replaced by input vacuum fluctuations at rate $\kappa$, which is much faster than the decoherence rate of the mechanical oscillator $\gamma$.

\subsection{Instantaneous multiple additions/subtractions} \label{sec:instantaneous_multiple_additions_subtractions}

In order to perform an instantaneous two-phonon addition/subtraction, we generalize the scheme with a single APD to a HBT-like scheme with two on-off detectors.
The leaking from the optomechanical cavity light (labeled "L") is sent to a balanced beamsplitter with a vacuum ancilla (labeled "A") in the other port.
At the two outputs of the beamsplitter, we place on-off detectors.
Each of the on-off detectors is characterized by a POVM consisting of two operators:
\begin{equation}
  \Pi_0^i = \dyad{ 0_i }{0 _i },
  \quad
  \Pi_1^i = \II^i - \Pi_0^i,
  \quad
  i = \text{A,L}.
\end{equation}
Here $\Pi_1^i$ corresponds to a detection in the arm labeled $i$, and $\Pi_0^i$ to no click.
Correspondingly, a coincidence count, when both detectors observe non-zero number of incoming photons, corresponds to the operator $\Pi_1\up{A}\Pi_1\up{L}$.
The system comprising mechanics with light and ancilla modes, is before the detection in the state $\rho\s{HBT}$.
The two clicks, in "L" and "A" modes, project the mechanical mode on the conditional state
\begin{equation}
  \rho\s{m}\up{HBT} = \frac{  \Tr\s{A,L} \mleft[ \Pi_1\up{L} \Pi_1\up{A} \rho\s{HBT} \mright] }{  \Tr\s{A,L,M} \mleft[ \Pi_1\up{L} \Pi_1\up{A} \rho\s{HBT} \mright] }.
\end{equation}
We omitted everywhere $\II\up{M}$, and will use shortened notation $\II\up{M} \otimes \II\up{L} \otimes \Pi_0\up{A} \mapsto \Pi_0\up{A}$ in what follows.
The state in square brackets can be written as
\newcommand*{\ppi}{\mathbb{\Pi}}
\begin{multline}
  \tilde \rho\s{m}\up{HBT} =
\mleft[ ( \II\up{L} - \Pi_0\up{L} ) \otimes ( \II\up{A} - \Pi_0\up{A} ) \rho\s{HBT} \mright] =
  \\
\mleft[
  \mleft(
  \II\up{LA} - \Pi_0\up{L} - \Pi_0\up{A} + \Pi_0\up{L} \Pi_0\up{A}
  \mright) \rho\s{HBT}
  \mright],
\end{multline}
For Gaussian states, this expression can be simplified further.

After the optomechanical interaction, the optomechanical system comprised of the mechanical mode and the leaking light, is in a Gaussian state described by the covariance matrix $\mmat V\s{ML}$.
The ancilla is in vacuum state with covariance matrix $\mmat V\s{A} = \II_2$, so that before the beamsplitter, the three-mode system is also in a Gaussian state with CM $\mmat V\s{MLA} = \mmat V\s{ML} \oplus \mmat V\s{A}$.
After the beamsplitter, the state remains Gaussian with the CM that reads
\begin{multline}
  \mmat V\s{HBT} = \mmat T\s{BS} \mmat V\s{MLA}  \mmat T\s{BS}^\intercal,
  \\
  \mmat T\s{BS} =
  \II_2 \oplus
  \begin{pmatrix}
    \sqrt{ T } \II_2 &  \sqrt{ 1 - T } \II_2
    \\
    - \sqrt{ 1 - T } \II_2 & \sqrt{ T } \II_2
  \end{pmatrix}.
\end{multline}
For a balanced beamsplitter, $T = 1/\sqrt{2}$.
Eventually, the state at the detector has a CM conveniently written in the block form
\begin{equation}
  \mmat V\s{HBT} =
  \begin{pmatrix}
    \mmat C\s{M,M}           & \mmat C\s{M,L}           & \mmat C \s{M,A} \\
    \mmat C\s{M,L}^\intercal & \mmat C\s{L,L}           & \mmat C \s{L,A} \\
    \mmat C\s{M,A}^\intercal & \mmat C\s{L,A}^\intercal & \mmat C \s{A,A}
  \end{pmatrix}
\end{equation}

Using the notation introduced in~\cref{sec:heralding_using_on_off_detectors}, for the numerator of the conditional state
\begin{align}
  \Tr\s{AL} [ \rho\s{HBT} ] & = \rho\s{G} ( \mmat C\s{M,M} );
  \\
  \Tr\s{AL} [ \Pi_0\up{L} \rho\s{HBT} ] & = p\s{c} ( \II\up{L} ) \Tr\s{A} \rho\s{c} ( \II\up{L} );
  \\
  \Tr\s{AL} [ \Pi_0\up{A} \rho\s{HBT} ] & = p\s{c} ( \II\up{A} ) \Tr\s{L} \rho\s{c} ( \II\up{A} );
  \\
  \Tr\s{AL} [ \Pi_0\up{A}\otimes \Pi_0\up{L} \rho\s{HBT} ] & = p\s{c} ( \II\up{AL} ) \rho\s{c} ( \II\up{AL} ),
\end{align}
where all $p\s{c}( \mmat O ) = p\s{c} (\mmat V\s{HBT}, \mmat O )$, and $\rho\s{c} (\mmat O) = \rho\s{c} (\mmat V\s{HBT}, \mmat O)$.

Tracing over certain modes of a Gaussian state simply removes corresponding rows and columns of a covariance matrix.
E.g. $\Tr\s{AL} \rG( \mmat V\s{HBT} ) = \rG( \mmat C\s{M,M})$.
For the probabilities in the denominator
\begin{align}
  \Tr\s{ALM} [ \rho\s{HBT} ] & = 1 ;
  \\
  \Tr\s{ALM} [ \Pi_0\up{L} \rho\s{HBT} ] & = p\s{c} ( \II\up{L} ) ;
  \\
  \Tr\s{ALM} [ \Pi_0\up{A} \rho\s{HBT} ] & = p\s{c} ( \II\up{A} ) ;
  \\
  \Tr\s{ALM} [ \Pi_0\up{A}\otimes \Pi_0\up{L} \rho\s{HBT} ] & = p\s{c} ( \II\up{AL} ).
\end{align}

The conditional state then reads
\begin{multline}
  \mleftright
  \rho\up{HBT}\s{m} =
  \left[ 1 - p\s{c} ( \eye\up{L}) - p\s{c} ( \eye\up{A}) + p\s{c} ( \eye\up{AL}) \right]^{-1}
  \\
  \times \bigg( \rG ( \mmat C\s{M,M} ) - p\s{c} ( \II\up{L} ) \Tr\s{A} \rho\s{c} ( \II\up{L} )
    \\
    - p\s{c} ( \II\up{A} ) \Tr\s{L} \rho\s{c} ( \II\up{A} ) + p\s{c} ( \II\up{AL} ) \rho\s{c} ( \II\up{AL} )
  \bigg).
\end{multline}
This is a negative sum of three Gaussian functions.

\section{Criteria of non-Gaussianity} \label{sec:criteria_of_non_gaussianity}

\subsection{Photon-number-based criteria} \label{sec:photon_number_based_criteria}

Using non-linear detectors such as on-off ones, and arrays thereof, can drive the initially thermal mechanical oscillations to non-Gaussian states.
By definition, these states are the ones whose Wigner functions are non-Gaussian.
Given a quantum state $\rho$ of an oscillator, its Wigner function is defined as
\begin{equation}
  \label{eq:wigner_def}
  W_\rho (x,y) = \frac{ 1 }{ \pi } \int_{-\infty}^{\infty} \ee^{ 2 \ii p y } \mel{ x - y }{\rho}{x + y } \dl{y}.
\end{equation}
Non-Gaussianity of this function can be whitnessed e.g. by a direct examination, given the full quantum state tomograpy, or by observing its negative values when the values of the function are known only in certain points of the phase space.
The methods relying on the reconstruction of the values of the Wigner function, however, can be demanding in an experiment.
Moreover, negativity of the Wigner function is quickly lost under loss which is a significant obstacle.
In practice, more precise methods of non-Gaussianity verification exist, in particular, based on the multiphoton contributions of the quantum states, that is, on the diagonal elements of the density matrix
\begin{equation}
  p_k = \mel{k}{\rho}{k}.
\end{equation}

In particular, on the plane $(p_0, p_1)$, only a region of this plane is accessible to Gaussian states~\cite{filip_detecting_2011,jezek_experimental_2011}.
The region can be parametrized as
\begin{equation}
  \label{eq:criteria:p0p1}
  p_0 = \frac{ \ee^{ - d^2 [ 1 - \tanh (r)] }}{ \cosh (r) },
  \quad
  p_1 = u \frac{ d^2 \ee^{ - d^2 [ 1 - \tanh(r) ] }}{\cosh^3 (r) },
\end{equation}
with $d^2 = ( \ee^{4r} - 1 )/4, \ 0 \leq u \leq 1, \ r \geq 0$.
This region is marked by filling in~\cref{fig:fig0}d.

A more fine-grained determination of non-Gaussianity can be performed having access to photon-number-resolving detection~\cite{lachman_faithful_2019,podhora_quantum_2022}.
In more detail, the $n-$photon non-Gaussianity is defined as the inability of the state to be represented as a result of an arbitrary Gaussian operation applied to an arbitrary combination of the $n-1$ lower Fock states.
This property is evaluated by finding the upper bounds for the $n-$photon contributions ($\braket{n}{\psi}$) of the corresponding states
\begin{equation}
  p_n\up{G} = \abs{\braket{n}{\psi_{n-1}} }^2 =  \abs{\bra{n} \hat S (r) \hat D (\alpha) \mleft[ \sum_{i = 0}^{n-1} c_i \ket{i} \mright]}^2,
\end{equation}
where $\hat D (\alpha) = \exp[ \alpha^* a - \alpha a\dg]$ is a displacement operator, and $\hat S(r) = \exp[ r (a\dg)^2 - r^* a^2 ]$ is a squeezing operator.
Observing a matrix element $\mel{n}{\rho}{n} > p\up{G}_n$ thus witnesses a $n-$photon non-Gaussianity of the state $\rho$.

\section{Quantifiers of non-Gaussian bunching capability} \label{sec:quantifiers_of_non_gaussianity}

In this section we derive the normalized Wigner function of the output state of a linear interferometric device with copies of our test state $\hat \rho$ entering all $n$ of its inputs.
In the extreme case of $n-1$ outputs being in vacuum, the (unnormalized) quantum state of the remaining output port is denoted $\tilde W(x,y)$.
Thanks to this state being rotationally-symmetric in the phase space, one can write
\begin{equation}
  \tilde W(x,y) = \tilde W( r \cos \phi , r \sin \phi ) = w\s{out}(r),
\end{equation}
where the expression for $w\s{out}$ reads~\cite{zapletal_multicopy_2017}
\begin{multline}
  \mleftright
  w\s{out} (r ) \propto
  \sum_{m_1 = 0}^M \sum_{m_2 = 0}^M \dots \sum_{m_n = 0}^M \left( \prod_{j=1}^n \frac{ p_{m_j}}{ m_j !}\right)
  \\
  \frac{ \left(\sum_{i=1}^n m_i\right)! }{ ( - n)^{\sum_{i=1}^n m_i} } \ee^{ - \frac{r^2}{2} } L_{\sum_{i=1}^n m_i } (r^2 ).
\end{multline}
Here $r^2 = x^2 + y^2$ is the distance from the phase-space point to the origin, $p_j = \mel{j}{\hat \rho}{j}$ are the multiphonon contributions of the state under examination, $L_s$ is the Laguerre polynomial of the order $s$.
To compute the norm, we note
\begin{equation}
  \iint_{-\infty}^{\infty} \tilde W (x ,y) \dl{x} \dl{y} = w_0 = 2 \pi \int_0^\infty r w(r) \dl{r}.
\end{equation}
Computing this last integral yields $w_0$ which then gives the \emph{normalized} Wigner function
\begin{equation}
  W(x,0) = \frac{ 2 \pi }{ w_0 } w\s{out}(x).
\end{equation}

Moreover, rotational symmetry of the Fock states and the states that we consider, allows a simplification of computation of the products of the form $\Tr[ \rho_1 \rho_2 ]$.
In particular, it is possible to write
\begin{multline}
  ( 4 \pi )^{-1} \Tr[ \rho_1 \rho_2 ] =
  \int_{-\infty}^{\infty} \dl{x} \dl{y} W_1 (x,y) W_2 (x,y)
  \\
  =
  \int_0^{\infty} \dl{r} \int_0^{2\pi} \dl{\phi} r  W_1 (r \cos \phi, r \sin \phi ) W_2 (r \cos \phi, r\sin \phi )
  \\
  =
  2 \pi \int_0^{\infty} \dl{r} r  w_1 (r ) w_2 (r )
\end{multline}

\section{Phonon-subtracted thermal states} \label{sec:phonon_subtracted_thermal_states}

\begin{figure}[htb]
  \centering
  \includegraphics[width=0.99\linewidth]{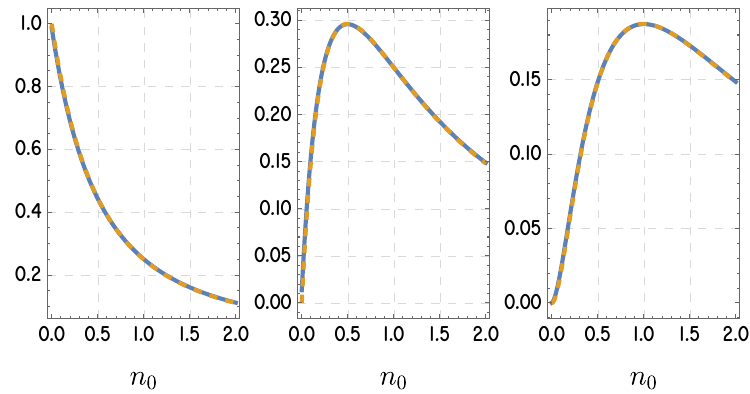}
  \caption{Multiphonon contributions of a single-phonon-subtracted thermal state of the mechanical oscillator: $\mel{k}{\rho}{k}$.
    From left to right, $k = 0,1,2$.
  Blue lines: simulation of the optomechanical subtraction,
yellow lines: ideal theoretical SPSTS.}
  \label{fig:spsts-phon-probs}
\end{figure}

Here we briefly elaborate on the process of phonon subtraction.
When the effective optomechanical interaction implements excitation hopping, detecting leaking photons heralds subtraction of a photon from the initial mechanical thermal state.
The heralded state approaches an ideal single-phonon-subtracted thermal state (SPSTS, up to normalization $\rho\s{SPSTS} \propto a\s{m} \rho\s{th} (n_0) a\s{m}\dg$).
In~\cref{fig:spsts-phon-probs} we compare the optomechanical subtraction with the ideal subtraction to find that the results are barely distinguishable.

For the ideal SPSTS, multiphonon probabilities equal
\begin{equation}
  p\s{SPSTS}{}_n (n_0) = (n+1) \frac{ n_0^n }{(n_0+1)^{n+2} } \leq 4 n^n \frac{(n+1) }{ (n+2)^{n+2} }.
\end{equation}
Unfortunately, despite being quantum non-Gaussian in the sense of criteria~\cref{eq:criteria:p0p1}, none of these probabilities exceed the multiphonon QNG thresholds.

\section{Detection of phase-randomized displacement} \label{sec:detection_of_phase_randomized_displacement}

\begin{figure}[htb]
  \centering
  \includegraphics[width=0.99\linewidth]{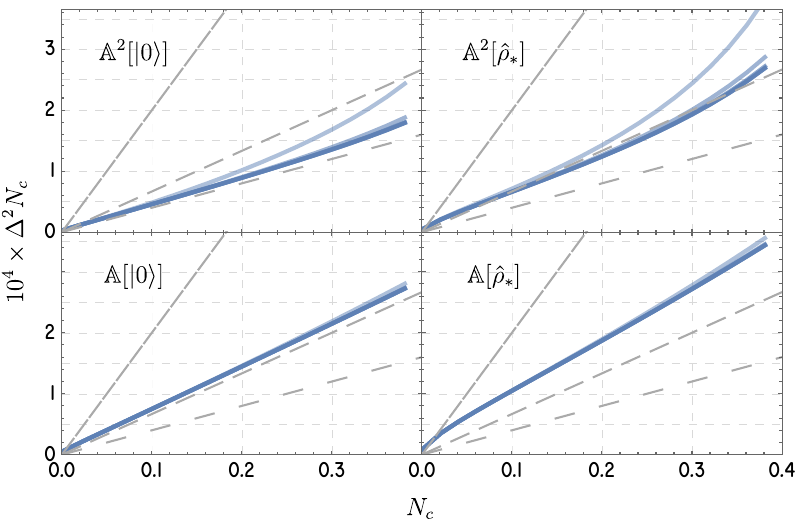}
  \caption{
    Impact of limited photon-number-resolving detectors on the estimation error of phase-randomized displacement.
    Different panels correspond to different states (notation same as in~\cref{fig:fisher_information}).
    Lines with different opacity correspond to different $k\s{max}$ in~\cref{eq:fishapp:finite}, from $k\s{max} = 2$ in the lightest curves, to $k\s{max} = 6$ in the darkest.
    From the graph follows that the lines with $k\s{max} \geq 3$ are indistinguishable.
  }
  \label{fig:fig_extra_fish}
\end{figure}

In this section we consider the effect of a detector that is capable of resolving only a finite number of phonons.
Such a detector is characterized by a POVM comprising operators
\(
\Pi_k = \dyad{k}{k},
\)
where $k$ takes integer values between $0$ and $k\s{max}$.

Accordingly,~\cref{eq:displ:fisher} is rewritten as
\begin{equation}
  \label{eq:fishapp:finite}
  F(N_c) = \sum_{n = 0}^{k\s{max}} \frac{ 1 }{ p\s{f} (n | N_c ) } \mleft( \frac{ \partial p\s{f} (n| N_c) }{ \partial N_c } \mright)^2 +
  \frac{ 1 }{ p_{k+}} \mleft( \frac{ \partial p_{k+} }{ \partial N_c } \mright)^2,
\end{equation}
where
\begin{equation}
  p_{k+} = \sum_{n = k\s{max} + 1 }^{ \infty } p\s{f} (n | N_c ).
\end{equation}

Result of our simulations for the same probe states as in the main text, are in~\cref{fig:fig_extra_fish}.
As is visible from the graph, after increasing $k\s{max} \geq 3$, the estimation error does not reduce further for the two-phonon added states, and $k\s{max} = 2$ is sufficient for single-phonon added states.

\bibliography{/home/omn/works/refs}
\end{document}